\documentstyle[amssymb,aps,preprint,psfig]{revtex}
\begin{document}
\title{On the violation of the Fermi-liquid picture in two-dimensional systems in
the presence of van Hove singularities}
\author{V.Yu.Irkhin$^{*}$ and A.A.Katanin}
\address{Institute of Metal Physics, 620219 Ekaterinburg, Russia}
\maketitle

\begin{abstract}
We consider the two-dimensional $t$-$t^{\prime }$ Hubbard model with the
Fermi level being close to the van Hove singularities. The phase diagram of
the model is discussed. In a broad energy region the self-energy at the
singularity points has a nearly-linear energy dependence. The corresponding
correction to the density of states is proportional to $\ln ^3|\varepsilon
|. $ Both real- and imaginary part of the self-energy increase near the
quantum phase transition into magnetically ordered or superconducting phase
which implies violation of the Fermi-liquid behavior. The application of the
results to cuprates is discussed.
\end{abstract}

\pacs{75.10.-b, 71.20.-b, 71.10.-w}

\section{Introduction}

Last decade, a possibility of a non-Fermi-liquid (NFL) behavior in
two-dimensional (2D) systems has been a subject of many theoretical
investigations. This question is especially interesting in connection with
the high-temperature superconductors where many unconventional features,
including pseudogap phenomena, are observed. Anderson \cite{Anderson} has
put forward the idea that a 2D system can demonstrate NFL behavior at
arbitrary small interelectron repulsion $U$ owing to a finite phase shift at
the Fermi energy. Even in the absence of such effects, a NFL state can occur
at small $U$ because of peculiarities of electron or spin and charge
fluctuation spectra.

In the usual 2D Fermi liquid the imaginary part of the self-energy (electron
damping) has the energy dependence ${\rm Im}\Sigma ({\bf k}_F,\varepsilon
)\propto \varepsilon ^2\ln |\varepsilon |$ \cite{Bloom,Fukuyama}, and the
temperature behavior of resistivity is $\rho \propto T^2\ln T$ \cite{Hod}.
However, these dependences do not describe experimental data on high-$T_c$
cooper-oxide compounds. To treat the anomalous behavior of these systems,
Varma et al \cite{MFL} proposed the phenomenological marginal Fermi-liquid
(MFL) theory where ${\rm Im}\Sigma ({\bf k}_F,\varepsilon )\propto
|\varepsilon |$. Then the electronic specific heat should demonstrate $T\ln
T $-dependence, and the resistivity the $T$-linear behavior. A similar
behavior in a broad temperature region can be obtained in the presence of
strong antiferromagnetic fluctuations for 2D and nested 3D systems \cite
{Mor,IK}. The linear energy dependence of the self-energy was also obtained
within the spin-fermion theory\cite{Chubukov,Chubukov1} which supposes that
the magnetic correlation length of the system is large enough.

Another explanation of anomalous electron properties of 2D lattice systems
can be found in the presence of the van Hove (VH) singularities \cite{MF}.
In this case, the bare electron density of states is logarithmically
divergent and to leading (second) order in $U$ we have the behavior ${\rm Im}%
\Sigma ({\bf k}_{VH},\varepsilon )\propto |\varepsilon |\ln |1/\varepsilon |$
at $|\varepsilon |\gg |\mu |$ (${\bf k}_{VH}$ is assumed to be VH point),
which differs slightly from the linear dependence in the MFL theory.
However, the important question occurs how this behavior changes with taking
into account higher-order terms. RPA calculations \cite{Onufr}, numerical
calculations within the FLEX approximation \cite{Kastr} and analytical
analysis \cite{Kastr} show that not too close to van Hove filling the
(quasi-) linear energy dependence of the self-energy takes place in a broad
range of the energies.

Within the renormalization-group approach, the behavior of the self-energy
was discussed in Refs. \cite{Dzy96,Menashe}. The conclusions of different
approaches turned out to be contradictory: a possibility of a NFL state is
discussed in Ref.\cite{Dzy96}, while a standard Fermi-liquid behavior was
found in Ref. \cite{Menashe}. It should be noted that in both the papers
only part of all the possible interaction channels was taken into account.
However, as it was discussed in Refs. \cite{OurVH,Salmhofer01}, all the
scattering channels are equally important, which can substantially change
the previous results.

In the present paper we consider the energy-dependence of the self-energy
with account of all the scattering channels. We demonstrate that a violation
of the Fermi-liquid picture takes place provided that the system is close to
quantum phase transition into a magnetically ordered or superconducting
state. Our approach can be considered as a generalization of that of Ref.
\cite{NestMFL} to the case of a non-nested Fermi surface. Note that our RG
analysis takes into account the contribution of the vicinity of VH points
only. For fillings close to VH one, this contribution is most singular and
vertices with the corresponding momenta are most divergent. This can be
checked by analysis of the results of complete RG approach which takes into
account the contribution of entire Fermi surface\cite{FurRice}. Unlike this
approach, however, present approach gives the possibility to take into
account all channels of electron scattering, as discussed in details in Ref.
\cite{OurVH}. Recent analysis \cite{Salmhofer01} which permits to consider
the interplay of different channels gives the results close to that obtained
earliar within two-patch and parquet equations approach \cite{OurVH}. Thus
we can expect that the approach based on consideration of the vicinities of
VH singularities is reliable and gives qualitatively correct results for
fillings close to VH one. The behavior of some physical properties and
application of the results obtained to cuprates is discussed in the
Conclusion.

\section{The model and quantum phase transitions}

We consider the $t$-$t^{\prime }$ Hubbard model on the square lattice:
\begin{equation}
H=\sum_{{\bf k}}\varepsilon _{{\bf k}}c_{{\bf k}\sigma }^{\dagger }c_{{\bf k}%
\sigma }+U\sum_in_{i\uparrow }n_{i\downarrow }  \label{H}
\end{equation}
with the electron spectrum
\begin{equation}
\varepsilon _{{\bf k}}=-2t(\cos k_x+\cos k_y)-4t^{\prime }(\cos k_x\cos
k_y+1)-\mu  \label{ek}
\end{equation}
Hereafter we assume $t>0,$ $t^{\prime }<0$ (which is the case for hole-doped
systems), $0\leq |t^{\prime }/t|<1/2$. For $t^{\prime }=0$ the Fermi surface
is nested, which results in peculiarities of physical properties \cite
{NestMFL,IK}. However, nesting is removed for $t^{\prime }/t\neq 0.$ For
arbitrary $t^{\prime }/t,$ the spectrum (\ref{ek}) contains VH singularities
connected with the points $A=(\pi ,0),$ $B=(0,\pi ).$ The chemical potential
$\mu $ is determined by the electron concentration $n$ and can be obtained
from the condition
\begin{equation}
n=\sum_{{\bf k}}f_{{\bf k}}
\end{equation}
with $f_{{\bf k}}=f(\varepsilon _{{\bf k}})$ being the Fermi function, so
that $\mu =0$ corresponds to Van Hove filling. Being expanded near the VH
singularity points, the spectrum\ (\ref{ek}) takes the form
\begin{mathletters}
\begin{eqnarray}
\varepsilon _{{\bf k}}^A &=&-2t(\sin ^2\varphi \overline{k}_x^2-\cos
^2\varphi k_y^2)-\mu  \label{eka} \\
\varepsilon _{{\bf k}}^B &=&2t(\cos ^2\varphi k_x^2-\sin ^2\varphi \overline{%
k}_y^2)-\mu  \label{ekb}
\end{eqnarray}
where $\overline{k}_x=\pi -k_x,$ $\overline{k}_y=\pi -k_y,\varphi $ is the
half of the angle between asymptotes at VH singularity, $2\varphi =\cos
^{-1}(-2t^{\prime }/t).$

We have a set of quantum phase transitions (QPT's) for the model (\ref{H}),
see Ref. \cite{OurVH}. We restrict ourselves to the case of small enough $%
U\lesssim 6t$, so that we may neglect the correlation effects connected with
Hubbard's subband formation (e.g., the Mott-Hubbard metal-insulator
transition). Provided that the electron concentration is not too close to
its van Hove value, i.e. the chemical potential satisfies $|\mu |>\mu _c$, $%
\mu _c$ being the critical value, we have the normal (paramagnetic and
non-superconducting) phase. When approaching the VH singularity, a QPT to
antiferromagnetic, superconducting or ferromagnetic state occurs. The type
of the ground state depends on $t^{\prime }/t$ and $U$. In particular, for $%
U=4t$ we have the antiferromagnetic ground state at $|t^{\prime }/t|<0.30$
and ferromagnetic one for larger $|t^{\prime }/t|$ \cite{OurVH}. The
critical electron concentrations for stability of these phases differ by
several percents from van Hove filling.

\section{Electron self-energy in the second order}

Consider first zero-temperature perturbative results. To second order the
expression for the electron self-energy has the form
\end{mathletters}
\begin{equation}
\Sigma ^{(2)}({\bf k},\varepsilon )=U^2\sum_{{\bf pq}}\frac{f_{{\bf p+q-k}%
}(1-f_{{\bf p}}-f_{{\bf q}})+f_{{\bf p}}f_{{\bf q}}}{\varepsilon
+\varepsilon _{{\bf p+q-k}}-\varepsilon _{{\bf p}}-\varepsilon _{{\bf q}}}
\label{Sigma}
\end{equation}
This contribution can be represented as a sum of three diagrams (Fig.2 a-c):
\begin{equation}
\Sigma ^{(2)}({\bf k},\varepsilon )=\Sigma _1({\bf k},\varepsilon )+\Sigma
_2({\bf k},\varepsilon )+\Sigma _3({\bf k},\varepsilon )
\end{equation}
When picking out the singularities we can put $\varepsilon _{{\bf p}%
}=\varepsilon _{{\bf p}}^A$ or $\varepsilon _{{\bf p}}=\varepsilon _{{\bf p}%
}^B$ for ${\bf p}$ being close to $A(B)$ van Hove points. Note that the term
$\Sigma _1$ was investigated earlier \cite{Dzy96,Menashe,Gonzalez}. We
restrict our consideration to the VH points at the Fermi surface, ${\bf k}%
_F=(0,\pi )$ or $(\pi ,0).$ The calculations at $|\varepsilon |\gg |\mu |$
yield:
\begin{eqnarray}
{\rm Re}\Sigma _1({\bf k}_F,\varepsilon ) &=&-(\ln 2)(g_0/\sin 2\varphi
)^2\varepsilon \ln ^2(\Lambda ^2|t/\varepsilon |)  \nonumber \\
{\rm Re}\Sigma _{2,3}({\bf k}_F,\varepsilon ) &=&-(g_0/\sin 2\varphi
)^2\varepsilon \left\{
\begin{array}{cc}
A_{2,3}\ln (\Lambda ^2|2t/\varepsilon |) & |\varepsilon |\ll 2t\cos 2\varphi
\\
k_{2,3}\ln 2\ln ^2(\Lambda ^2|2t/\varepsilon |) & |\varepsilon |\gg 2t\cos
2\varphi
\end{array}
\right.
\end{eqnarray}
where $g_0=U/(4\pi ^2t)$ is the dimensionless coupling constant, $\Lambda
\sim 1$ is the ultraviolet momentum cutoff, $k_2=1,$ $k_3=2/3,$%
\begin{mathletters}
\label{A23}
\begin{eqnarray}
A_2 &=&\int\limits_0^{\cos 2\varphi }\frac{dx}x\ln \frac{\varepsilon _x-x}{%
\varepsilon _x}+\frac 12\int\limits_{\cos 2\varphi }^{1/\cos 2\varphi }\frac{%
dx}x\ln \frac{\varepsilon _x+x}{\varepsilon _x}  \nonumber \\
&&\ \ \ \ \ \ \ \ \ \ \ \ \ \ \ \ \ \ \ +\frac 12\int\limits_{-\infty }^0%
\frac{dx}x\ln \frac{\varepsilon _x}{\varepsilon _x+x} \\
A_3 &=&\int\limits_0^{\cos 2\varphi }\frac{dx}{2\varepsilon _x+x}\ln \left( -%
\frac x{2\varepsilon _x}\right) +\frac 12\int\limits_{\cos 2\varphi
}^{1/\cos 2\varphi }\frac{dx}{2\varepsilon _x+x}\ln \frac{2(\varepsilon _x+x)%
}x  \nonumber \\
&&\ \ \ \ \ \ \ \ \ \ \ \ \ \ \ \ \ \ \ +\frac 12\int\limits_{-\infty }^0%
\frac{dx}{2\varepsilon _x+x}\ln \frac x{2(\varepsilon _x+x)}
\end{eqnarray}
with $\varepsilon _x=(x-\cos 2\varphi )(1-x\cos 2\varphi )/\sin ^22\varphi $%
. For small energies $|\varepsilon |\ll |\mu |$ the real part remains linear
in energy with the difference that logarithmical divergences are cut at $%
|\mu |$ rather than at $|\varepsilon |$. The corresponding imaginary parts
at $|\varepsilon |\gg |\mu |$ read
\end{mathletters}
\begin{eqnarray}
{\rm Im}\Sigma _1({\bf k}_F,\varepsilon ) &=&-(\pi \ln 2)(g_0/\sin 2\varphi
)^2|\varepsilon |\ln (\Lambda ^2|t/\varepsilon |)  \nonumber \\
{\rm Im}\Sigma _{2,3}({\bf k}_F,\varepsilon ) &=&-\pi (g_0/\sin 2\varphi
)^2|\varepsilon |\left\{
\begin{array}{cc}
B_{2,3} & |\varepsilon |\ll 2t\cos 2\varphi \\
k_{2,3}\ln 2\ln (\Lambda ^2|t/\varepsilon |) & |\varepsilon |\gg 2t\cos
2\varphi
\end{array}
\right.
\end{eqnarray}
where
\begin{eqnarray*}
B_2 &=&(A_{21}+A_{23})\theta (\varepsilon )+A_{22}\theta (-\varepsilon ) \\
B_3 &=&(A_{31}+A_{33})\theta (-\varepsilon )+A_{32}\theta (\varepsilon )
\end{eqnarray*}
and $A_{2n}$ and $A_{3n}$ are $n$-th summands in the definition of $A_2$ and
$A_3,$ Eq. (\ref{A23}), respectively, $\theta (\varepsilon )$ is the step
function. At $t^{\prime }/t\neq 0$ we have ${\rm Im}\Sigma ({\bf k}%
_F,\varepsilon )\neq {\rm Im}\Sigma ({\bf k}_F,-\varepsilon )$ because of
the absence of the particle-hole symmetry for the less singular terms $%
\Sigma _{2,3}({\bf k}_F,\varepsilon )$. This fact results in an asymmetry of
the electron density of states near the Fermi level and can be important for
some physical properties, e.g., thermoelectric power. Unlike the real part
of the self-energy, the imaginary part changes its dependence to quadratic
one at $|\varepsilon |\ll |\mu |$ demonstrating a conventional Fermi-liquid
behavior in this region.

\section{Renormalization}

Both real and imaginary parts of the self-energy contain large logarithms at
$|\mu |\ll |\varepsilon |\ll t$. Therefore we can introduce the logarithmic
variable $\lambda =\ln (\Lambda |t/\varepsilon |^{1/2}).$ Then the leading
terms of the expansion in the powers of interaction strength can be written
down as
\[
\Sigma (\lambda )=A\lambda g_0^2\varepsilon (1+C_1g_0\lambda
+D_1g_0^2\lambda +...)
\]
To perform the summation of leading logarithms in the self-energy, we
introduce the vertices $\gamma _i(\lambda ),$ $i=1,...4$ (Fig. 2), and
consider the renormalization of $\Sigma $. As discussed in Refs. \cite
{Led,Fur,OurVH}, $\gamma _i(\lambda )$ can be determined from the
renormalization-group (RG) equations
\begin{eqnarray}
\gamma _1^{\prime } &=&2d_1(\lambda )\gamma _1(\gamma _2-\gamma
_1)+2d_2\gamma _1\gamma _4-2\,d_3\gamma _1\gamma _2  \nonumber \\
\gamma _2^{\prime } &=&d_1(\lambda )(\gamma _2^2+\gamma _3^2)+2d_2(\gamma
_1-\gamma _2)\gamma _4-d_3(\gamma _1^2+\gamma _2^2)  \nonumber \\
\gamma _3^{\prime } &=&-2d_0(\lambda )\gamma _3\gamma _4+2d_1(\lambda
)\gamma _3(2\gamma _2-\gamma _1)  \nonumber \\
\gamma _4^{\prime } &=&-d_0(\lambda )(\gamma _3^2+\gamma _4^2)+d_2(\gamma
_1^2+2\gamma _1\gamma _2-2\gamma _2^2+\gamma _4^2)  \label{TwoPatch}
\end{eqnarray}
where $\gamma _i^{\prime }\equiv d\gamma _i/d\lambda $,
\begin{eqnarray}
d_0(\lambda ) &=&2c_0\lambda ;  \nonumber \\
d_1(\lambda ) &=&2\left\{
\begin{array}{cc}
\lambda , & \lambda <2z_{{\bf Q}} \\
z_{{\bf Q}}, & \lambda >2z_{{\bf Q}}
\end{array}
\right.  \nonumber \\
d_2 &=&2z_0;\;d_3=2c_{{\bf Q}}
\end{eqnarray}
The quantities
\[
z_0=c_0=1/\sin (2\varphi )=1/\sqrt{1-R^2}
\]
are the prelogarithmic factors in small-momentum particle-hole and
particle-particle bubble, while
\begin{eqnarray}
z_{{\bf Q}} &=&\ln [(1+\sqrt{1-R^2})/R]  \nonumber \\
c_{{\bf Q}} &=&\tan ^{-1}(R/\sqrt{1-R^2})/R  \label{zc}
\end{eqnarray}
are the prelogarithmic factors in particle-hole and particle-particle bubble
with momenta close to ${\bf Q}=(\pi {\bf ,}\pi ){\bf ,}$ $R=-2t^{\prime }/t$%
. Equations (\ref{TwoPatch}) should be solved with the initial condition $%
\gamma _i(0)=g_0$. Because of the presence of double-logarithmic terms, the
corresponding RG equations are only approximate. However, the comparison of
the results of their solution \cite{OurVH} with the parquet approach \cite
{OurVH} and the RG approach which takes into account the contribution of the
whole Fermi-surface \cite{Salmhofer01} shows that they reproduce well the
renormalization of the couplings.

The magnetic or superconducting instabilities manifest in the divergence of
the vertices $\gamma _i(\lambda )$ at some critical scale $\lambda _c.$ This
is connected with the critical energy scales discussed in Sect. II as $\mu
_c\sim T_c\sim \Lambda \exp (-2\lambda _c),$ for a detailed discussion see
Ref.\cite{OurVH}. (In the absence of interlayer coupling the quantity $T_c$
has the meaning of a temperature of crossover into the state with pronounced
short-range order (or pseudogap state) rather than of a phase transition
temperature.) For $\lambda $ close to $\lambda _c$ the solution of the Eqs. (%
\ref{TwoPatch}) can be represented in the form
\begin{equation}
\gamma _i=\frac{\gamma _i^c}{\lambda _c-\lambda }  \label{gc}
\end{equation}
The corresponding correlation length is given by
\begin{equation}
\xi ^{-2}=C_\xi |\mu |(\lambda _c-\lambda )/t  \label{cr}
\end{equation}
where $C_\xi $ is the universal number. For $T\ll |\mu |$ we should stop the
scaling at
\begin{equation}
\lambda ^{*}=\ln [\Lambda t^{1/2}/\max (|\mu |,|\varepsilon |)^{1/2}]
\end{equation}
while $T\gg |\mu |$ at
\begin{equation}
\lambda ^{*}=\ln [\Lambda t^{1/2}/\max (T,|\varepsilon |)^{1/2}]
\end{equation}
The condition $|\mu |>\mu _c\ $or $T>T_c$ guarantees that $\lambda
^{*}<\lambda _c,$ i.e. the system is not ordered and RG approach is
applicable.

Now we consider the renormalization of the self-energy. We follow the method
of Refs.\cite{Solyom,Bourbonnais,Dupuis}. Defining the scale-dependent
quasiparticle residue
\[
Z(\lambda )=Z_1(\lambda )Z_2(\lambda )Z_3(\lambda )
\]
where $Z_i$ ($i=1,2,3$) is the contribution of $i$-th diagram, we have the
RG equations
\begin{eqnarray}
\frac{d\ln Z_1(\lambda )}{d\lambda } &=&-(8\ln 2)\lambda \gamma _4^2/\sin
^22\varphi  \nonumber \\
\frac{d\ln Z_2(\lambda )}{d\lambda } &=&-D_2(\lambda )(\gamma _1^2+\gamma
_2^2-\gamma _1\gamma _2)/\sin ^22\varphi  \nonumber \\
\frac{d\ln Z_3(\lambda )}{d\lambda } &=&-D_3(\lambda )\gamma _3^2/\sin
^22\varphi  \label{ZZ}
\end{eqnarray}
Here
\[
D_{2,3}(\lambda )=\left\{
\begin{array}{cc}
4k_{2,3}\lambda \ln 2 & \lambda <(1/2)\ln (1/\cos 2\varphi ) \\
A_{2,3} & \lambda >(1/2)\ln (1/\cos 2\varphi )
\end{array}
\right.
\]
and the summation over spin indices in vertices is performed. Then the real
part of the self-energy can be found as
\begin{equation}
\text{Re}\Sigma ({\bf k}_F,\varepsilon )=\varepsilon \ln Z(\lambda ^{*})
\label{SigmaZ}
\end{equation}
(see, e.g., Ref. \cite{Dupuis}). After calculating Re$\Sigma $, the
imaginary part of the self-energy can be obtained from the Kramers-Kronig
relations.

\section{Results of calculations}

First, we demonstrate our approach in a simple case with the only non-zero
vertex, $\gamma _4\neq 0.$ As discussed in Ref. \cite{OurVH}, this case
corresponds to $t^{\prime }\rightarrow -t/2,$ the ground state at van Hove
filling being ferromagnetic (flat-band ferromagnetism). Then we have
\begin{equation}
\gamma _4=\frac{g_0}{1+g_0(c_0\lambda ^2-2z_0\lambda )}  \label{g4}
\end{equation}
The vertex (\ref{g4}) diverges in the critical point
\begin{equation}
\lambda _c=\frac 12\ln \frac{\Lambda ^2t}{\max (\mu ,T)}=1-\sqrt{1-1/(z_0g_0)%
}
\end{equation}
where we have put $z_0=c_0$ in $\lambda _c$ according to (\ref{zc}). From (%
\ref{ZZ}) we obtain $Z_2=Z_3=1$ and
\begin{equation}
Z_1=\exp \left[ -\frac{4\ln 2}{\sin ^22\varphi }\frac{g_0^2\lambda (\lambda
-1)}{(1-g\,_0z_0)\,(1+\,g\,_0z_0\lambda ^2-2g_0z_0\,\lambda )}\right]
\end{equation}
For $\lambda $ being close to $\lambda _c$ we have
\begin{equation}
\gamma _4\sim \xi ^2(T=0)\sim (\lambda _c-\lambda )^{-1}  \label{Cr}
\end{equation}

As well as in the nesting case (see, e.g., Ref. \cite{NestMFL}), the
quasiparticle weight $Z$ vanishes exponentially in the QPT point. Note that
this vanishing is much faster than the inverse-logarithmic dependence
\begin{equation}
Z\sim \frac 1{\ln ^q(\lambda _c-\lambda )},\;q\simeq 0.35
\end{equation}
obtained in Ref. \cite{Dzy96} where only one scattering channel was taken
into account. Besides nearly-linear dependences in $\varepsilon $ (with
logarithmic corrections), both real and imaginary part of the self-energy
contain large $\varepsilon $-dependent factors of the order of $1/(\lambda
_c-\lambda )^2\propto \xi ^4(0),$ which occur because of the divergence of
the vertex (\ref{g4}) at $\lambda \rightarrow $ $\lambda _c$.

As discussed in Ref. \cite{OurVH}, for $|t^{\prime }/t|$ being not very
close to $1/2$ we have an interplay of all the scattering channels, so that
we have to solve Eqs. (\ref{TwoPatch}), (\ref{ZZ}) numerically. We also
calculate the quasiparticle spectral weight in VH points of the Fermi
surface,
\begin{equation}
A({\bf k}_{VH},\varepsilon )=-\frac 1\pi \frac{{\rm Im}\Sigma ({\bf k}%
_{VH},\varepsilon )}{[\varepsilon +\mu -{\rm Re}\Sigma ({\bf k}%
_{VH},\varepsilon )]^2+[{\rm Im}\Sigma ({\bf k}_{VH},\varepsilon )]^2}
\end{equation}
The results of the calculations for $t^{\prime }=-0.45t,$ $\mu /t=0.3,$ $T=0$
(nearly ferromagnetic ground state, $\mu _c/t=0.04$) are shown in Fig. 3. As
well as in the second order of the perturbation theory, at $|\varepsilon
|\ll |\mu |$ we have the 2D behavior of self-energy
\begin{equation}
\text{Re}\Sigma ({\bf k}_F,\varepsilon )\propto \varepsilon ,\;\text{Im}%
\Sigma ({\bf k}_F,\varepsilon )\propto \varepsilon ^2,  \label{S1}
\end{equation}
while $|\mu |\ll |\varepsilon |\ll t$ it changes to van Hove behavior,
\begin{equation}
\text{Re}\Sigma ({\bf k}_F,\varepsilon )\propto \varepsilon \ln ^2(\Lambda
t/\varepsilon ),\;\text{Im}\Sigma ({\bf k}_F,\varepsilon )\propto
|\varepsilon |\ln (\Lambda t/\varepsilon )  \label{S2}
\end{equation}
Finally, at $\varepsilon =\varepsilon _0\sim t$ the real part of self-energy
has a maximum and then decreases with farther increasing of $\varepsilon ,$
while imaginary part is almost a constant at $\varepsilon \sim t.$ The slope
of RG results in the region $|\mu |\ll |\varepsilon |\ll t$ is substantially
higher that the results of the second-order perturbation theory, the
imaginary part of the self-energy at $|\varepsilon |\geq |\mu |$ also
becomes large enough. The peak in the quasiparticle weight obtained within
RG approach is more broad than in the second-order perturbation theory,
which is the consequence of larger electron damping in the letter case. With
the increasing the chemical potential (so that the system is moved away from
QPT), the higher-order renormalizations become less important and the
behavior of the self-energy reproduces the result of the second-order
perturbation theory.

The imaginary part of self-energy at finite temperature and $\varepsilon =0$
(the inverse quasiparticle lifetime at the Fermi surface) can be obtained
from the scaling arguments,
\begin{equation}
\gamma (T)=-[{\rm Im}\Sigma ({\bf k}_F,\varepsilon )|_{T=0}]_{\varepsilon
\rightarrow T}
\end{equation}
Thus it is given by the same Figs. 3b with the replacement $\varepsilon
\rightarrow T$ and also demonstrates the linear behavior in some temperature
region.

In the above consideration we neglected completely the damping of the
particle-particle and particle-hole excitations as well as other
non-singular contributions. These non-singular terms can be neglected
provided that the condition
\begin{equation}
\xi ^{-2}\gg \max (|\mu |,|\varepsilon |)/t
\end{equation}
is satisfied, i.e. not too close to quantum phase transition into the
corresponding ordered state and at not too high energies. Close to the
quantum phase transition into the ordered state, the peaks in the
self-energy should become asymmetric, as discussed in Ref. \cite{Onufr}, and
the pseudogap can be formed. On the other hand, in the limit of large enough
$|\mu |,$ i.e. at
\begin{equation}
\max (|\varepsilon |/t,\xi ^{-2})\ll |\mu |/t
\end{equation}
the spin-fermion theory \cite{Chubukov1} is applicable.

\section{Discussion and conclusions}

In the present paper we have investigated within a scaling approach the
energy dependence of the real and imaginary part of the electron self-energy
in the presence of van Hove singularities. We have restricted ourselves to
the regions of the $\mu -T$ phase diagram, where the Fermi level is not too
close to VH points (disordered ground state, $\mu _c<|\mu |\ll t$ with $\mu
_c$ is the critical chemical potential) or temperature is above the critical
value ($T>T_c$ with $T_c\sim \mu _c$).

Provided the system is not too close to QPT (at $\mu _c<|\mu |\ll t$), the
imaginary part of the self-energy at VH points demonstrates in a broad
energy region $|\mu |\ll |\varepsilon |\ll t$ a nearly-linear behavior, $%
{\rm Im}\Sigma ({\bf k}_F,\varepsilon )\propto |\varepsilon |\ln
(t/|\varepsilon |)$, which is close to that in the marginal Fermi-liquid
concept \cite{MFL} (however, in MFL the linear behavior of ${\rm Im}\Sigma (%
{\bf k}_F,\varepsilon )$ takes place for arbitrary ${\bf k}_F$). The real
part of the self-energy behaves as $\varepsilon \ln ^2(t/|\varepsilon |).$
The linear energy dependence of the self-energy (\ref{S2}) and the behaviour
of the slope with changing $\mu $ are in agreement with the RG results of
Ref. \cite{Honerkamp} which takes into account of the contribution of the
whole Fermi-surface. These dependences are also similar to those obtained
within the spin-fermion model for a nearly antiferromagnetic state \cite
{Chubukov,Chubukov1}, although in our case they have a different nature and
are governed by van Hove singularities themselves rather than by closeness
to antiferromagnetic quantum phase transition. The role of a characteristic
spin-fluctuation frequency $\omega _{sf},$ which separates the Fermi-liquid
and MFL regimes, belongs in our case to the chemical potential $|\mu |$ ($%
\mu =0$ corresponds to VH filling). Another difference is that in the
presence of VH singularities the linear dependence of self-energy takes
place already in the weak-coupling regime.

Near QPT, the renormalizations become important because of the large
ground-state correlation length which enters renormalized vertices.
Therefore both real and imaginary parts of the self-energy increase
considerably as $\xi ^2(T=0).$ Such an anomalous behavior also implies a
strong violation of the Fermi-liquid and even MFL picture. Note that the
anomalies under consideration may induce the electron topological transition
with the truncation of the Fermi surface (see, e.g. Ref. \cite{Chubukov2}).
The RG approach used is not able to describe the magnetically ordered or
superconducting state. By this reason, the renormalized-classical regime $%
T<T_c,$ $|\mu |<\mu _c$ should be considered within other approaches, see,
e.g., Ref. \cite{Vilk}.

In 2D situation, the physical properities near QPT should demonstrate
singularities which are stronger than those in the 3D case. The correction
to the electron density of states $N(\varepsilon )$ reads
\begin{equation}
\delta N(\varepsilon )=-\sum_{{\bf k}\sigma }\left[ {\rm Re}\Sigma ({\bf k,}%
\varepsilon )\delta ^{\prime }(\varepsilon -\varepsilon _{{\bf k}})+\frac 1%
\pi {\rm Im}\Sigma ({\bf k,}\varepsilon )/(\varepsilon -\varepsilon _{{\bf k}%
})^2\right]
\end{equation}
Main contribution to the integral comes from the vicinity of VH points where
the bare density of states is logarithmically divergent. Taking into account
that ${\rm Re}\Sigma ^{(2)}({\bf k},\varepsilon )\propto \varepsilon \ln
^2\max \{|\varepsilon |,|\varepsilon _{{\bf k}}|\}$, we obtain for the first
(coherent) term in the square brackets which originates from renormalization
of quasiparticle spectrum
\begin{equation}
\delta N_{coh}(\varepsilon )\propto \ln ^3(t/|\varepsilon |),\varepsilon \gg
\mu .
\end{equation}
The calculation of the second (incoherent, non-quasiparticle) term requires
the full form of $\Sigma ^{(2)}({\bf k,}\varepsilon ),$ Eq.(\ref{Sigma}),
and leads to the result
\begin{equation}
\delta N_{incoh}(\varepsilon )\propto \ln ^2(t/|\varepsilon +\mu |).
\end{equation}
Although this divergence is slightly weaker than of the coherent term, it is
not cut at $\varepsilon =-\mu .$ Thereby the bare VH singularity becomes
considerably enhanced. Note that the divergence of the density of states
together with its asymmetry in $\varepsilon $ may lead to peculiarities of
thermoelectric power owing to impurity scattering, cf. Ref.\cite{IK}.

To leading (second) order the contribution to electronic specific heat owing
to VH singularities has the form $\delta C\propto T\ln ^3(t/\max \{|\mu
|,T\})$. The resistivity (inverse transport relaxation time) should
demonstrate at $T>\mu $ the behavior $\rho \propto T\ln ^2(t/T).$ The
calculations are similar to those of Ref.\cite{IK} for the antiferromagnetic
state, extra logarithmic factors coming from VH singularities. The crossover
from quadratic to nearly linear temperature dependence of resistivity is
confirmed by experimental data for cuprates (see, e.g., the results of Ref.%
\cite{Ono} for the LaSrCuO system).

Thus the divergences in the many-electron system with VH singularities are
stronger than those in the MFL theory. Near QPT, we can expect that all the
physical properties are strongly renormalized, the renormalizations being
dependent on the type of the ordered phase. This problem will be considered
elsewhere.

Finally we consider the application of the results obtained to cuprate
systems. A nearly linear energy dependence of ${\rm Im}\Sigma ({\bf k}%
_F,\varepsilon )$ at VH points, which is similar to our results, was
observed for the system Bi2212 in ARPES experiments \cite{Valla}. For the
system La$_{2-x}$Sr$_x$CuO$_4$ with the doping $x_c\simeq 0.2$ (which is
slightly larger than the optimal one) the Fermi surface crosses VH points
\cite{LaFS}. The density of states \cite{LaAIPES}, specific heat coefficient
and Pauli susceptibility \cite{LaChi} substantially grow near this doping.
The mass enhancement factor $m^{*}/m$ demonstrates a similar behavior. The
additional experimental investigations of the self-energy near $(\pi ,0)$
point of La$_{2-x}$Sr$_x$CuO$_4$ would be interesting in this respect.

\section{Acknowledgments}

We are grateful to Andrey Chubukov for useful discussions of the physical
picture and experimental situation for cuprates. The research described was
supported in part by Grant No.00-15-96544 from the Russian Basic Research
Foundation (Support of Scientific Schools).

\newpage
{\sc Figures}

\psfig{file=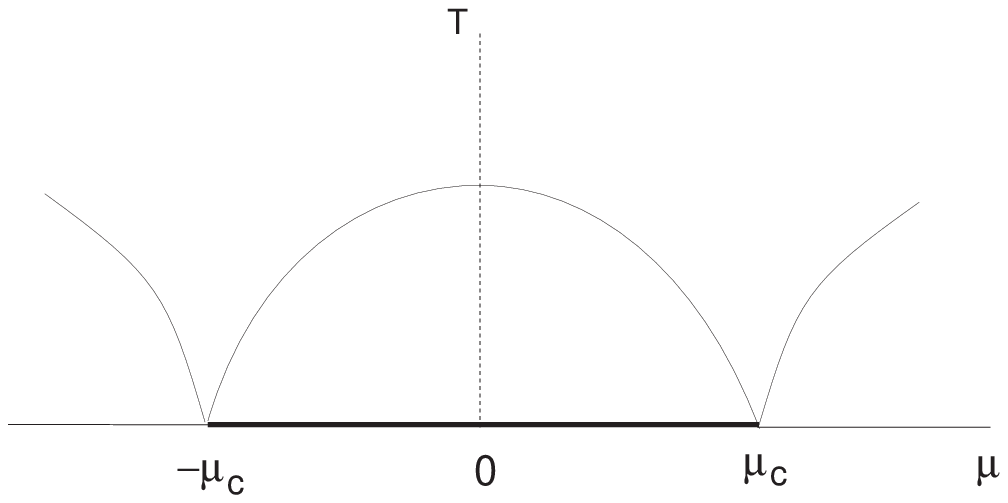}

Fig.1. A qualitative $T-\mu $ phase diagram in the vicinity of quantum phase
transitions. The chemical potential $\mu $ is referred to the Van Hove
singularity. Bold line denotes the ordered ground state.

\vspace{1.0in}

\psfig{file=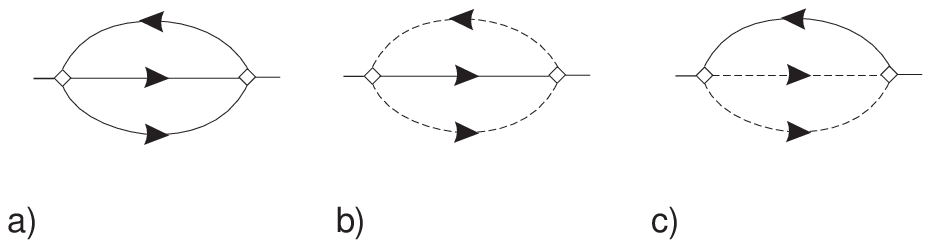}

Fig.2. The second-order diagrams for the electron self-energy. Solid and
dashed lines correspond to electrons with momenta close to ($0,\pi $) and ($%
\pi ,0$) van Hove singularities respectively

\newpage

\psfig{file=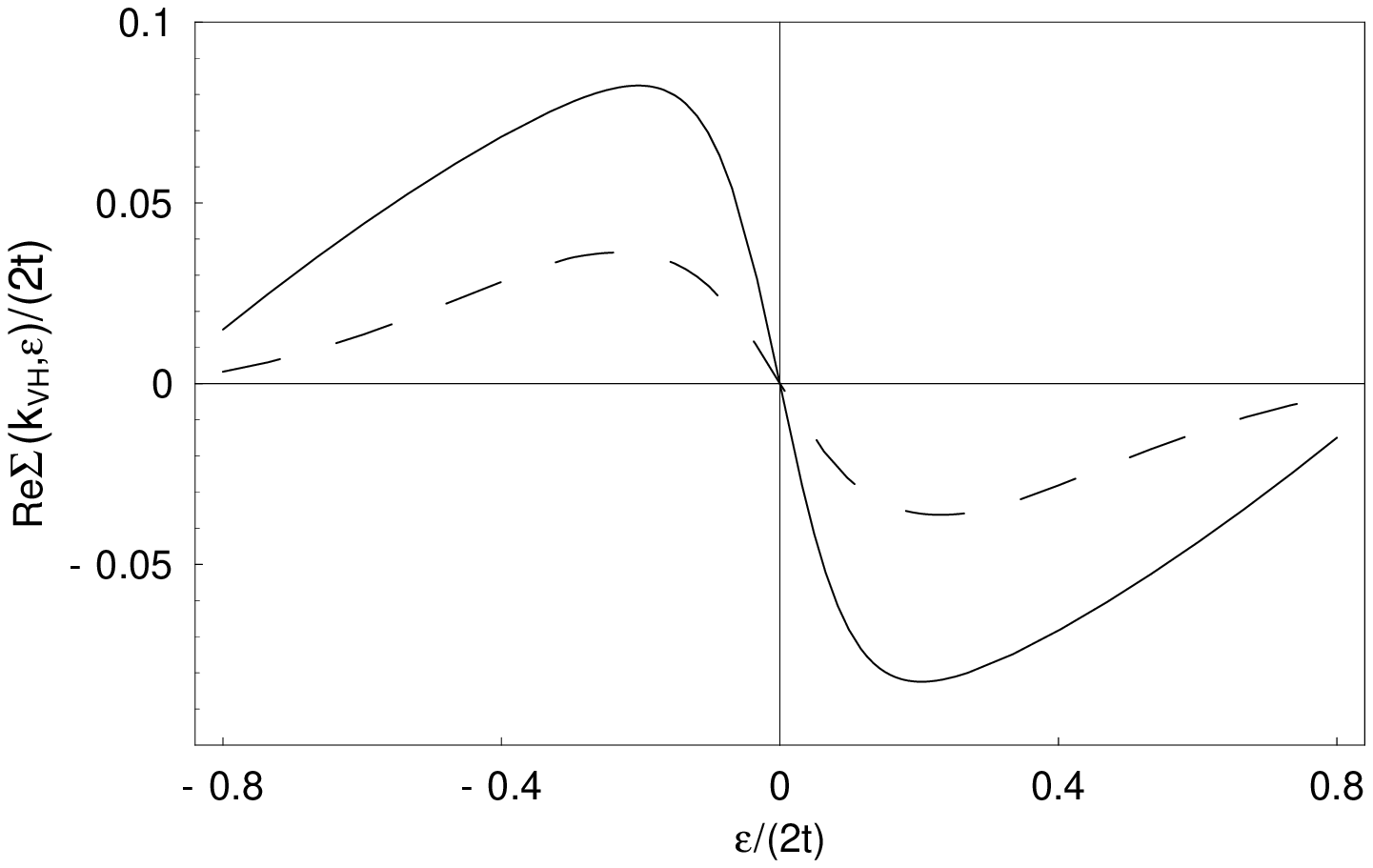}

\vspace{0.1in}

\psfig{file=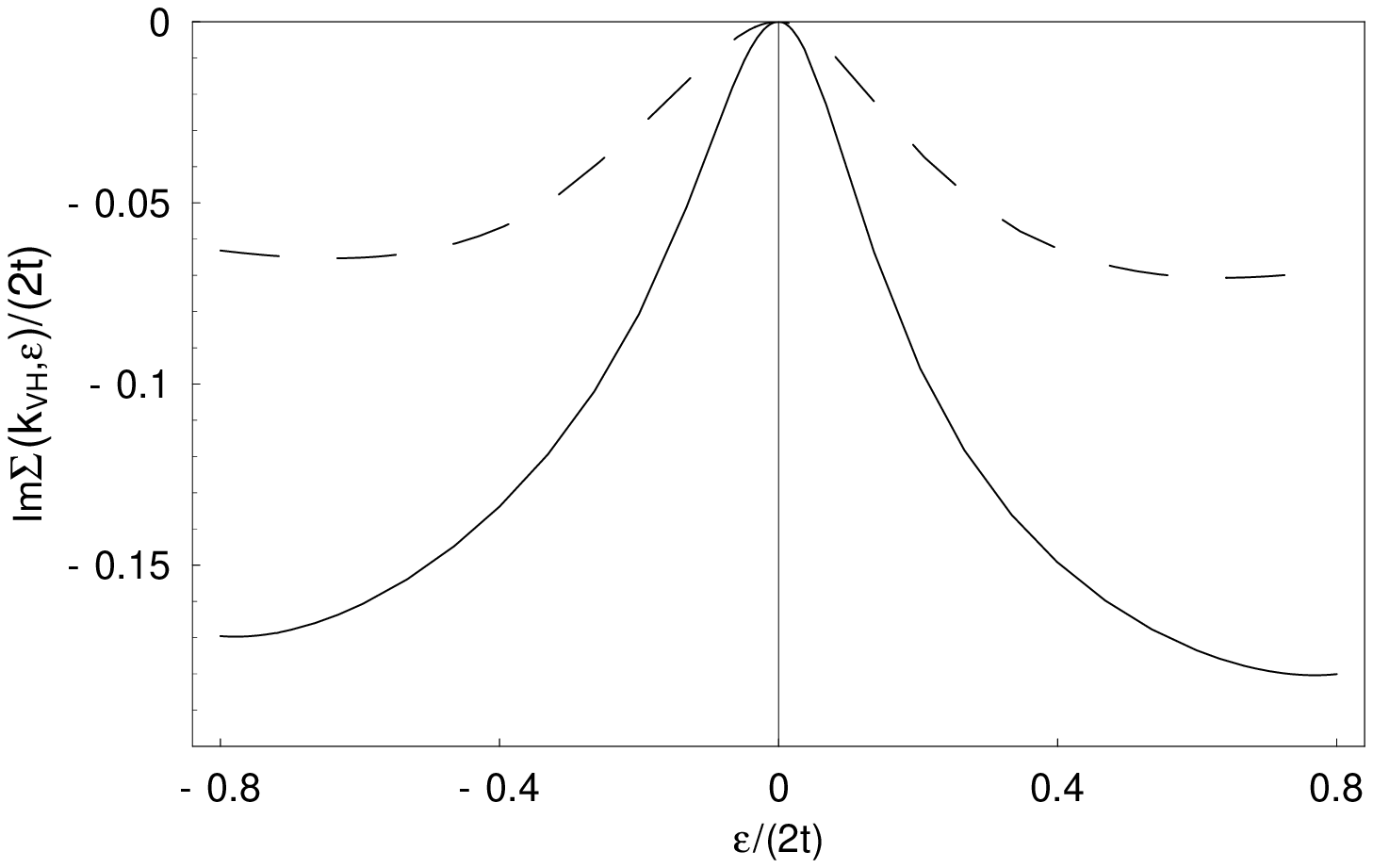} \newpage

\psfig{file=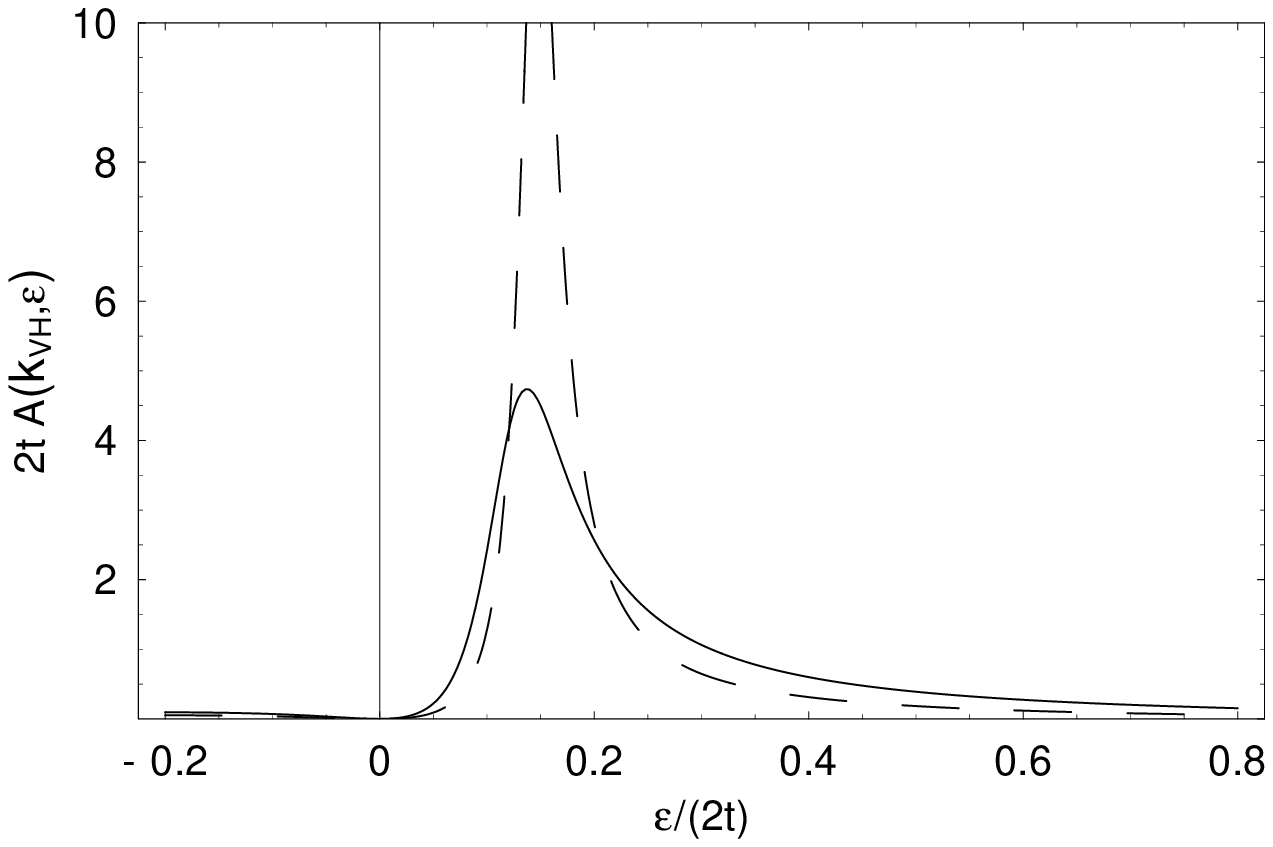}

Fig.3. The energy dependences of real (a) and imaginary (b) parts of the
self-energy at VH points, and the spectral weight\ (c) for $t^{\prime
}/t=-0.45,$ $U=4t,$ $\mu /(2t)=0.15\ $and $T=0.$ The dashed line corresponds
to the second-order perturbation result.

\end{document}